\documentclass[prd,twocolumn,floatfix,superscriptaddress,showpacs]{revtex4-2}
\pdfoutput=1
 \usepackage[utf8]{inputenc}
\usepackage{graphicx}
\usepackage{epsfig}
\usepackage{bm}
\usepackage{amssymb}
\usepackage{float}
\usepackage{amsmath}
\usepackage{dcolumn}
\usepackage{cancel}
\usepackage[colorlinks]{hyperref}
\usepackage[usenames,dvipsnames]{color}
\hypersetup{
     breaklinks=true,
    pdfstartview={FitH},    
    colorlinks=true,       
    linkcolor=blue,          
    citecolor=red,        
    filecolor=magenta,      
    urlcolor=blue,           
    anchorcolor=green,      
    linktocpage=true
}

\usepackage{amsmath,latexsym}
\usepackage{placeins}
\usepackage[toc,page]{appendix}
\newcommand{\df}{\dot{\phi}}
\newcommand{\MP}{M_{Pl}^2}

\newcommand{\egr}{\epsilon_{GR}}
\newcommand{\egena}{\epsilon_{G1}}
\newcommand{\egdio}{\epsilon_{G2}}
\newcommand{\egtria}{\epsilon_{G3}}
\newcommand{\egtessera}{\epsilon_{G4}}
\newcommand{\eniena}{\epsilon_{N1}}
\newcommand{\enidio}{\epsilon_{N2}}
\newcommand{\enitria}{\epsilon_{N3}}
\newcommand{\enitessera}{\epsilon_{N4}}
\begin{document}

\title{Successful
Higgs inflation from combined  nonminimal and derivative couplings
  }

\author{Stelios Karydas }
\email{stkarydas@mail.ntua.gr} \affiliation{Physics Division,
National Technical University of Athens, 15780 Zografou Campus,
Athens, Greece.}

\author{Eleftherios Papantonopoulos}
\email{lpapa@central.ntua.gr} \affiliation{Physics Division,
National Technical University of Athens, 15780 Zografou Campus,
Athens, Greece.}

\author{Emmanuel N. Saridakis}
\email{msaridak@noa.gr} \affiliation{National Observatory of Athens, Lofos Nymfon, 11852 Athens, Greece}
\affiliation{Department of Astronomy, School of Physical Sciences,
University of Science and Technology of China, Hefei 230026, People's Republic of China}

\begin{abstract}
\noindent{
We propose an  inflationary scenario based on the concurrent presence of
non-minimal coupling (NMC) and generalized non-minimal derivative coupling
(GNMDC), in the context of Higgs inflation. The combined construction maintains
the advantages of the individual scenarios without sharing their disadvantages.
In particular, a long inflationary phase can be easily achieved due to the
gravitational friction effect owed to the GNMDC, without leading to
trans-Planckian values and unitarity violation. Additionally, the
tensor-to-scalar ratio remains to low values due to the NMC contribution.
Finally, the instabilities related to the squared sound-speed of scalar
perturbations, which plague the simple GNMDC scenarios, are now healed
due to the domination of the NMC contribution and the damping of the GNMDC
effects during the reheating era. These features make  scenarios
with  nonminimal and derivative couplings to gravity  successful
candidates for the description of inflation.
}
\end{abstract}
\maketitle

\section{Introduction}\label{intro}

The inflationary scenario according  to which an early exponential expansion of
the Universe takes place, offers a compelling explanation for the initial
conditions of a hot Big Bang (for reviews see \cite{Olive:1989nu,
Lyth:1998xn,Martin:2013tda}). This inflationary
description of the early phases
of the Universe can be viewed as the effect of the dynamics of a scalar field
called inflaton. At the same time, observations based on the Cosmic Microwave
Background Radiation  (CMBR) offer increasingly precise constraints to test the
inflationary paradigm, as well as the theory of gravity that operates at very
high densities. Moreover, there has been a significant effort regarding the formation
of primordial black holes, during a super slow-roll phase during the inflationary period, which could in fact
be a viable Dark Matter
candidate (see \cite{Khlopov:2008myu,Chongchitnan:2006wx,Chen:2019zza,
Garcia-Bellido:2017mdw,Cai:2018tuh,Ballesteros:2018wlw,Germani:2018jgr,
Fu:2019ttf,Lu:2019sti,Dalianis:2018ymb,Yi:2020cut,Yi:2020kmq,Pi:2017}). Hence, the
physics surrounding
inflation is of particular significance to various aspects of our understanding
of the Universe, and as such  early universe cosmology provides the grounds to
test
and choose between a significant number of inflationary models. To identify a
viable one, one should study the dynamics of the full system of the inflaton
field and gravity.

In an attempt to describe the early cosmological evolution according to recent
observational results, gravity theories that are based on modifications of
Einstein Gravity were proposed. Two of the most common ways to modify the
standard Theory of Relativity, are introducing higher order curvature terms,
and/or including scalar fields that are nonminimally coupled to gravity.
Higher-order corrections to the Einstein-Hilbert action arise naturally in the
gravitational effective action of String Theory \cite{Gross:1986mw}. On the
other hand, introducing extra scalar fields, which are non-minimally coupled to
gravity, is a thoroughly studied way to modify the standard theory of General
Relativity (GR) and results to what is known as scalar-tensor theory
\cite{Fujii:2003pa}.  A particularly well-studied scalar-tensor theory is the
one obtained through the Horndeski Lagrangian \cite{Horndeski:1974wa}. These
theories yield field equations of second order and hence they do not produce
ghost instabilities \cite{Ostrogradsky:1850fid}. Moreover, many scalar-tensor
theories share a classical Galilean symmetry
\cite{Nicolis:2008in,Deffayet:2009wt,Deffayet:2009mn,Deffayet:2011gsz,
Kobayashi:2011nu,Harko:2016xip,Kamada:2010qe,Saridakis:2021lqd}.

One simple subclass of Horndeski theories  is obtained with the use of a scalar
field coupled to the Ricci scalar, which is known as Non-Minimal Coupling
(NMC). Such a construction goes beyond  the simple
case of GR plus a scalar field and thus it can improve the inflationary
behavior. In particular, by taking a
NMC of the form $\xi \phi^2$, if the scale $\xi$ is large enough the resulting
inflationary phase is long enough
\cite{Salopek:1989,Fakir:1990,Kaiser:1994,Komatsu:1999,Nozari:2008,Ren:2014}. In fact,
it was shown that a rather well-behaved phenomenology is obtained, since the
tensor
to scalar ratio is particularly low, and easily inside the Planck 2018
observational limits. Additionally, there have also been other works that have
utilized a
different NMC \cite{Park:2008} or a matrix configuration for the inflaton field \cite{Ashoorioon:2019}. However, although NMC models with large
coupling values are very efficient in producing improved inflationary
phenomenology, large  coupling constants
lead to problems related to the unitarity of this theory, and thus  are
not desirable from a quantum mechanical perspective
\cite{Bezrukov:2007,Barvinsky:2008,Bezrukov:2008,Bellido:2008,Simone:2008,
Bezrukov:2008b,Burgess:2009,Barbon:2009,
Bezrukov:2009,Barvinsky:2009,Clark:2009, Barvinsky:2009b,
 Einhorn:2009,Lerner:2009b,Burgess:2010,Mazumdar:2010,
Geng:2017mic,Fumagalli:2016},
if one is to have a single field model. A different picture is obtained when
multi field theories are studied, and it has been argued \cite{Hertzberg:2010,
Giudice:2010} that in such theories these problems do not exist. Moreover,
other attempts without unitarity issues have been made in a similar context,
utilizing a Palatini formulation of gravity
\cite{Tenkanen:2017,McDonald:2020,Tenkanen:2020}, or by taking into account
additional interactions \cite{Lerner:2010b,Lola:2020lvk}.

On the other hand, in Horndeski theory one of the
most well-studied terms is the one corresponding to the non-minimal derivative
coupling (NMDC) of the scalar field to the Einstein tensor. This term has
interesting implications both on small and large scales for black hole
physics  \cite{Kolyvaris:2011fk, Rinaldi:2012vy,
Kolyvaris:2013zfa, Babichev:2013cya, Cisterna:2014nua,  Charmousis:2014zaa, Koutsoumbas:2015ekk, Anabalon:2013, Cisterna:2015, Cisterna:2016}, for dark energy
\cite{Saridakis:2010mf, Dent:2013awa} and for inflation
\cite{Amendola:1993uh,Sheikhahmadi:2016wyz} respectively. For a recent review,
see \cite{Papantonopoulos:2019eff}. Concerning inflation, the main advantage of
NMDC is that it is free from unitarity problems, and this led to
the established model of new Higgs inflation
\cite{Germani:2015plv}.

As it has been shown, the non-minimal derivative coupling acts as a
friction mechanism, and therefore from an inflationary model-building point of
view  it allows for the implementation of a slow-roll phase
\cite{Amendola:1993uh,Sushkov:2009hk},  as well as for inflation with
potentials such as the Standard-Model Higgs to be realized
\cite{Germani:2010gm}. In light of the above, it becomes a very attractive term within the framework of Horndeski theory.
Moreover, such models are consistently described within supergravity
\cite{Farakos:2012je, Farakos:2013zya} via the gauge kinematic function
\cite{Dalianis:2014sqa}.
An extensive study of the NMDC predictions is performed in
\cite{Dalianis:2016wpu}, where the dynamics of both the inflationary slow-roll
phase and the reheating phase were considered. In particular, the NMDC
oscillations of the inflaton are very rapid and remain undamped for a very
lengthy period \cite{Sadjadi:2012zp, Ghalee:2013ada, Yang:2015pga,
Gumjudpai:2015vio, Yi:2016jqr, Gialamas:2020vto, Myung:2016twf, Ema:2015oaa, Ema:2016hlw}, affecting heavy particle production
\cite{Koutsoumbas:2013boa}. However, such oscillations, where the NMDC remains
dominant over the standard GR term, are problematic in terms of stability of
the post-inflationary system. This is due to the oscillations of the
sound-speed squared between positive and negative values \cite{Ema:2015oaa},
implying that scalar fluctuations are exponentially enhanced.

To avoid this instability, the non-minimal kinetic term must  cease to be the
dominant (or even co-leading) term, when compared to the canonical kinetic
term.
However, if this condition is  to be met, the model effectively reduces to that
of a canonical scalar field in GR  even during the slow-roll period, and the
advantages of the NMDC are lost. Nevertheless,  one can generalize the NMDC
term, since it is a special case of the Horndeski Lagrangian density, and
consider Lagrangians of the form
\cite{Deffayet:2011gsz,Kobayashi:2011nu,Harko:2016xip}
\begin{align} \label{lagran}
{\cal L}_5=G_5(\phi, X) G^{\mu\nu}\partial_\mu \partial_\nu \phi \,~,
\end{align}
where $X=-\partial_\mu \phi \partial^\mu \phi /2$.
If the function $G_5(\phi, X)$ is chosen to be,  $G_5(\phi,
X)=-\phi/(2M^2)$, one gets the simplest NMDC possible, since after integration by parts the derivative coupling becomes constant. This however leads
to the problematic post-inflationary evolution.
Instead, in \cite{Dalianis:2019vit} it was shown that if  one chooses a more
general function $G_5(\phi, X)=G(\phi)\, \xi(X)$, the phenomenology of the
Horndeski terms becomes richer, both during inflation and reheating stages.

In the case where $G(\phi)\propto \phi$ this Generalized NMDC term (GNMDC)
essentially vanishes when the inflaton field approaches the minimum of the
potential. Thus,  the system,  after a few oscillations, transits to the
dynamics of a canonical kinetic term in GR, leading to a more manageable and
reliable behavior, dominated by GR dynamics during the reheating stage. In
fact
it was shown that with this kind of term the inflationary phenomenology
generated in a Higgs potential was in very good agreement with
observations. Furthermore, the tight bounds on the speed of Gravitational
Waves (GWs) extracted by recent observations
\cite{Abbott:2016blz,Abbott:2016nmj,Baker:2017hug} and from the solar system constraints \cite{Gonzalez:2020vzl}, were dismissive of the NMDC
\cite{Ezquiaga:2017ekz,Gong:2017kim}, since an NMDC term playing the role of
dark energy can produce superluminal tensor perturbations
\cite{Germani:2010gm,Germani:2011ua} in Friedmann-Lemaitre-Robertson-Walker cosmological backgrounds and also.  A GNMDC
of the form $G(\phi)\propto \phi$  however can heal  this problem, since after
the end of slow roll inflation it has essentially decoupled from the dynamics
of the system since it becomes negligible. However, it was also shown that the 
sound speed square was not
completely healed of the oscillations between positive and negative values,
albeit this problem was significantly ameliorated. One then, would have to seek
for further modifications that could entirely heal the theories that are
modified with non-canonical kinetic terms of this form, from the sound-speed
related instabilities, and possibly even further improve the observables of
inflation.

The motivation of this work is based on the above discussion, according to
which  neither   the NMC nor
NMDC scenarios are completely free of disadvantages and problems when a
desirable phenomenology is achieved. Hence, we are interested in
investigating  a simple combination of the NMC and GNMDC terms,
that  can alleviate the   problems
of both of these standardized modifications. In particular, the GNMDC's
gravitational friction effect allows for the $\xi$ and $\phi_*$ to be lowered
enough to not violate unitarity, while the late time domination of the NMC term
ensures that no sound-speed related instabilities occur. Moreover, a lowering
of
the tensor-to-scalar ratio of this combined theory is obtained as compared to
the GNMDC case and a desirable theory is achieved.

This manuscript is organized as follows.   In Section \ref{standalones} we
briefly analyze some basic results of each of the NMC and GNMDC terms as
standalone modifications of GR. In Section \ref{combined},  we present the
combined scenario of inflation in the presence  of the NMC and GNMDC terms.
Then in Section \ref{specificstandalones} we proceed to a detailed numerical
investigation in a Higgs potential for a variety of interesting cases, with the purpose of demonstrating the general results obtained in Section \ref{combined}, thus clearly showing the advantages of this scenario. Finally, in Section \ref{conclusions} we summarize our results.

\section{Nonminimal coupling and  generalized nonminimal derivative coupling
as standalone modifications }\label{standalones}

In this Section we present a short synopsis of inflationary models resulting
from  general relativity plus nonminimal coupling  (GR+NMC)
and from     general relativity plus generalized nonminimal derivative coupling
 (GR+GNMDC), which have been studied extensively in the
literature.

In studying inflationary models it is  of great importance to perturbatively
study the effects of inflation, since every inflationary model provides a rich
phenomenology related to scalar and tensorial perturbations. This phenomenology
sets the observational testing grounds for all inflationary models.
Specifically, in order to test their viability, one
needs to compare the predictions of a variety of quantities with their
corresponding observed values, mainly obtained through CMBR. These observable quantities,   include the power
spectrum of the scalar perturbations, $\mathcal{P_R}$, the scalar spectral
index (tilt) $n_s$, and the tensor-to-scalar ratio $r$, while a specific
amount of e-folds is also required in order for  the
horizon and flatness problems to be efficiently  solved. In Appendix
\ref{perturbations} we include a short review of the usual steps taken in this
direction. A full analysis of single-field perturbations (without 
soft-properties considerations \cite{Saridakis:2021qxb}) has been performed in
a number of works, e.g. in \cite{Ema:2015oaa,Tsujikawa:2012mk,Kobayashi:2019} 
.
%
%
%
%
%
\subsection{Inflation with nonminimal coupling}

The action of this modification to GR is written in the form
\begin{align}\label{NMCaction}
	S=\int d^4x \sqrt{-g} \left[ f(\phi) R-\frac{\partial_\mu \phi \partial^\mu
\phi}{2}-V(\phi)\right],
\end{align}
and the most studied coupling of this form in the literature is
$f(\phi)=\xi \phi^2$. Nonminimal coupling (NMC) as a standalone modification to
GR, when taking the
form $f(\phi)=\xi \phi^2$ in a Higgs potential, has been shown to produce
remarkably low
tensor-to-scalar ratio values. Additionally, it  has no post-inflationary
instability issues,
since $c_s^2$ can be shown to be identically equal to 1, regardless of the form
of the NMC. Nevertheless, it was shown that it does not preserve unitarity and
thus it is problematic from a quantum-mechanical point of view, since the
combination $\xi \phi^2$
takes values larger than $M_{Pl}$ in order to yield a long enough inflation
\cite{Bezrukov:2007,Barvinsky:2008,Bezrukov:2008,Bellido:2008,Simone:2008,
Bezrukov:2008b,Burgess:2009,Barbon:2009,
Bezrukov:2009,Barvinsky:2009,Clark:2009, Barvinsky:2009b,
 Einhorn:2009,Lerner:2009b,Burgess:2010,Mazumdar:2010,
Geng:2017mic}.

We consider a homogeneous and isotropic
 flat   Friedmann-Robertson-Walker (FRW) geometry with
  metric
\begin{equation}
\label{FRWmetric}
ds^{2}=-dt^{2}+a^{2}(t)\delta_{ij}dx^{i}dx^{j}\,,
\end{equation}
where $a(t)$ is the scale factor.
The Friedmann equations of this scenario are
\begin{align}\label{Fried00NMC}
	3\MP H^2=V(\phi)+\frac{\df^2}{2}-6\xi \left[f(\phi) H^2+f'(\phi) \df
H\right],
\end{align}

\begin{align}
\label{Fried11NMC}
 \MP	( 2 \dot{H}+3   H^2)=
-\xi  \left[2 \df^2 f''(\phi)+4 H \df f'(\phi)+2
\ddot{\phi} f'(\phi) \right. \nonumber \\
+4
\left.f(\phi) \dot{H}+6 H^2 f(\phi)\right]-\frac{\df^2}{2}+V(\phi),
\end{align}
and the scalar-field equation of motion reads as:
\begin{align}\label{KGinNMC}
	\ddot{\phi}+3 H \df- 6 \xi f'(\phi) \left(\dot{H}+2 H^2\right)+V'(\phi)=0~.
\end{align}
However, in order to calculate the inflationary observables,  the convenient approach  is
to perform a conformal transformation, thus passing to the
Einstein frame.
By choosing
$
	\hat{g}_{\mu\nu}=\Omega^2(x) g_{\mu\nu}$,
with
\[
	\Omega^2(x)=\frac{16\pi}{\MP} f(\phi)~,
\]
and defining a new scalar field $\varphi$ and potential $U$ such that
\[
	\frac{d\varphi}{d\phi}\equiv
\sqrt{\frac{\MP}{8\pi}\frac{f(\phi)+3f'^2(\phi)}{2f^2(\phi)}}~,
\]
\[
	U(\varphi)\equiv \Omega^{-4} V(\phi)~,
\]
then the action is brought to the Einstein-frame equivalent form
\begin{align}
 	S=\int d^4x\sqrt{-\hat{g}}\left[\frac{\MP}{2}\hat{R}-\frac{\partial_\mu
\varphi \partial^\mu \varphi}{2}-U(\varphi) \right]~,
\end{align}
where the quantities in the Einstein frame are denoted with a hat.

It has been shown that to first order  one can write the spectral index and
tensor-to-scalar ratio as \cite{Martin:2013tda}
\begin{align}
	1-n_s=6 \epsilon_U-2\delta_U~,
\end{align}
\begin{align}
	r=16\epsilon_U~,
\end{align}
where  we have defined the slow-roll parameters
\begin{eqnarray}
	&&\epsilon_U  =\frac{\MP}{2}\left(\frac{U'}{U}\right)~,\\
	&&
\delta_U=\MP\frac{U''}{U}~.
\end{eqnarray}
Moreover, it can be easily shown that for an arbitrary coupling $f(\phi)$  the
$c_s^2$ of the GR+NMC scenario is identically equal to 1, by simple replacement
of the above equations into equation \eqref{sspddef}.

We mention here that in the Einstein frame  the potential $U$ is essentially
flat for large values of
the NMC term ($\xi \phi^2 \gg M_{Pl}$), hence the field rolls very slowly and
the slow-roll parameters  $\epsilon_U$ and $\delta_U$ are very small,
yielding a
correspondingly small $r$.
This last conclusion is what entails one of the basic results of single field,
NMC, Higgs inflation with the coupling form $f(\phi)=\phi^2$. Nevertheless,
as we mentioned above, these particularly attractive features of a very low $r$
and a long inflation, come  at the cost
of $\xi \phi^2 > M_{Pl}$, leading to non-unitarity. In order to solve this
problem one should consider other couplings of the scalar field to gravity, as
the one described in the next subsection.

%
%
%
%
%
%
\subsection{Inflation with nonminimal derivative coupling}

We now turn to the scenario according to which the generalized nonminimal
derivative coupling is a stand-alone
modification to GR. As we discussed in
the Introduction   in the general framework of Horndeski theories  nonminimal
derivative coupling (NMDC) holds a
particular position, due to its attractive feature of ``gravitational
friction'', i.e. the phenomenon according to which a single inflaton field,
when rolling down a potential, stays in ``slow roll'' for a significantly
lengthier period   compared to GR. This results to an easier realization of
inflation and a richer phenomenology, studied extensively in the literature
\cite{Germani:2010gm,Sadjadi:2012zp, Ghalee:2013ada,
Gumjudpai:2015vio,Gialamas:2020vto,
Myung:2016twf, Ema:2015oaa, Ema:2016hlw, Fumagalli:2017}.

However, among other effects   it has been argued that a standalone
NMDC modification to GR creates post-inflationary instabilities,
due to the fact that the NMDC term
remains dominant after the slow-roll period and this may lead to $c_s^2<0$. As a
result, a further intuitive modification,
dubbed generalized nonminimal
derivative coupling (GNMDC)  was proposed in
\cite{Harko:2016xip,Dalianis:2019vit}, where it was shown that
when the derivative coupling with the Einstein tensor is of the form
$G(\phi)\partial_\mu\phi \partial_\nu\phi G^{\mu\nu}$, then this problem is
significantly ameliorated.
In particular, the action of this modification to GR can be written in the form
\begin{align}\label{GNMDCaction}
	S=\int d^4x \sqrt{-g} \left[\frac{\MP}{2} R+G_5(\phi, X)
G^{\mu\nu}\partial_\mu
\partial_\nu \phi  -V(\phi)\right]~,
\end{align}
where $G^{\mu\nu}$ is the Einstein tensor. Hence, by considering only a
$\phi$-dependence of the $G_5$ function, the Friedmann equations of
this scenario are \cite{Harko:2016xip,Dalianis:2019vit}
\begin{align}\label{Fried00GNMDC}
&&	3 \MP H^2=9 H^2 G(\phi) \df^2+\frac{1}{2} \df^2+V(\phi)~,
\end{align}
\begin{align}
\label{Fried11GNMDC}
\MP	( 2 \dot{H}+3   H^2)&=V(\phi)-\frac{\df^2}{2}
+2 H \df^3 G'(\phi)\nonumber \\
&+G(\phi) \left(2 \dot{H} \df^2+3 H^2 \df^2+4 H \df
\ddot{\phi}\right)~, 	
\end{align}
while the scalar-field equation of motion reads as
\begin{align}\label{KGinGNMDC}
	\ddot{\phi}&+V'(\phi)+3 H \df+3 H^2 \df^2 G'(\phi)\nonumber \\
	&+G(\phi) \left(12 H H' \df+6 H^2 \ddot{\phi}+18 H^3
\df\right)=0~.
\end{align}
Note that the function $G(\phi)$  results from  $G_5$, after integrating by
parts, namely $G(\phi)=-G'_5(\phi)$.

In the class of models that  include a non-canonical
kinetic term, the gravitational friction effect  offers the
ground for very efficient inflationary predictions,
since the slow-roll
conditions can be   easily satisfied. In particular, in order to investigate
 inflation in the slow-roll approximation
  we define the slow-roll parameters
	\begin{equation}
\begin{aligned}[t]
	\epsilon =-\frac{\dot{H}}{H^2}~,
\end{aligned}
\qquad
\begin{aligned}[t]
	\delta =
\frac{\ddot{\phi}}{H\dot{\phi}}~,
\end{aligned}
\end{equation}

\begin{equation}
\begin{aligned}[t]
	\epsilon_V =\frac{\MP}{2}\left(\frac{V'}{V}\right)^2~,
\end{aligned}
\qquad
\begin{aligned}[t]
	\eta_V
\equiv
\frac{\MP}{2}\frac{V''}{V}~.
\end{aligned}
\end{equation}

The slow-roll approximation  holds when $\epsilon \ll 1$ and $\delta \ll 1$,
and thus $\dot{H} \ll H^2$
and $\ddot{\phi}\ll 3H\dot{\phi}$, and in this case the Friedmann equations
\eqref{Fried00GNMDC}, \eqref{KGinGNMDC} are simplified to
\begin{align} \label{fried00inGRNMDCSRA}
	3 \MP H^2\approx V(\phi)~,
\end{align}
\begin{align} \label{KGinGRNMDCSRA}
	3H\df \left[1 +6G(\phi)H^2+G'(\phi) H \df \right] +V'(\phi) \approx 0~.
\end{align}
Hence, under the slow-roll approximations, the first slow-roll
parameter, $\epsilon$, can then be written in the form
\begin{align} \label{epsilonbroken}
\epsilon \approx    \epsilon_\text{GR}+\epsilon_D+\epsilon_{\cal{B}}~,
\end{align}
where
\begin{equation}
\begin{aligned}[t]
	  \epsilon_D\equiv  \frac{3G(\phi) \df^2}{\MP}~,
\end{aligned}
\qquad
\begin{aligned}[t]
	  \epsilon_{\cal B}\equiv   \frac{\df^2}{  \MP H^2}   G'(\phi) H
\df~,
\end{aligned}
\end{equation}
These two functions correspond to $\egena$ and $\egtessera$ of equation \eqref{auxilepsilons} that we will later use. Moreover,
\begin{align}
	\epsilon_\text{GR}\equiv \frac{ \dot{\phi}^2}{2  \MP H^2}~,
\end{align}
where the quantity  $\epsilon_\text{GR}$ corresponds to the result of the GR
case, while   $\epsilon_D$   is the leading term during slow-roll.

The
GNMDC term has the effect of decreasing  the $\epsilon$ parameter and hence increases the
slow-roll period.
In fact, in the slow-roll approximation, equation
\eqref{epsilonbroken} can be brought to the form
\begin{align} \label{epsilonwithcal}
\epsilon=\epsilon_V \frac{\mathcal{A}-
2\mathcal{B}}{\left(\mathcal{A}+\mathcal{B}\right)^2}~,
\end{align}
with  ${\cal A } \equiv1+6H^2 G(\phi)$ and $ {\cal B} \equiv G'(\phi)H\df$.
Additionally,   the squared sound speed of the
scalar perturbations is found to be  \cite{Dalianis:2019vit}
\begin{align}\label{cssqGRNMDC}
	c_s^2= &\left[1-\frac{\epsilon_D}{3}   +6H^2 G(\phi)
 (1+\epsilon_D) \right]^{-1}\cdot \nonumber \\
\cdot&\Bigl\{1+\epsilon_D
+6H^2 G(\phi)\left[  1+ \epsilon_D+\frac{4
\epsilon_D(1-\epsilon_D)}{3(3-\epsilon_D)}
 \right]  \Bigr.\nonumber \\
&\Bigl.\text{ } +12{\dot{H}} G(\phi) \left({1-\frac{\epsilon_D}{3}}\right)
\Bigr\}~.
\end{align}
Furthermore,   we can approximate the number of
e-folds as \cite{Dalianis:2019vit}
\begin{align}
N \approx \frac{1}{M_{Pl}} \int_{\phi_{end}}^{\phi}
\frac{\mathcal{A}+\mathcal{B}}{\sqrt{2\epsilon_V}} d\phi~.
\end{align}
As one can see,   for $G(\phi)\rightarrow 0$ all the above expressions restore
the canonical case.
Finally, concerning the inflationary observables, the
scalar power spectrum  can be brought to the form  \cite{Dalianis:2019vit}
\begin{align} \label{PsinGRNMDC}
\mathcal{P}_\mathcal{R}=\frac{H^2}{8 \pi^2 \MP \epsilon_V}
\left[\mathcal{A} + \mathcal{B}
+\mathcal{O}\left(\frac{\mathcal{B}^2}{\mathcal{A}}\right)  \right]~,
\end{align}
the scalar spectral index becomes
\begin{align}\label{nsinGRNMDC}
	1-n_s\approx 8\epsilon-2\eta+\epsilon M_{Pl}
\frac{G'(\phi)}{G(\phi)}\sqrt{\frac{2}{\epsilon_V}}~,
\end{align}
with $\eta\equiv \frac{\eta_V}{\mathcal{A}}$, whilst the tensor-to-scalar ratio
is written as
\begin{align}\label{rinGRNMDC}
	r=16\frac{\epsilon_V}{\mathcal{A}+\mathcal{B}}~.
\end{align}

Let us now consider a specific model of GNMDC. In particular, we will focus on
the case
\begin{align}\label{Fphiform}
	G(\phi)=\frac{\alpha
\phi^{\alpha-1}}{2M^{\alpha+1}},
\end{align}
 which recovers the simple
NMDC for  $\alpha=1$
(see  \cite{Dalianis:2019vit} for the different case of
$G(\phi)=\frac{e^{\tau\phi/M_{Pl}}}{M^2}$).
Within the framework of this particular modification,
it  can be shown that as $\alpha$ becomes larger then the post-inflationary
instabilities
related to $c_s^2<0$ become significantly shorter as compared to the simple
NMDC $(\alpha=1)$.
  This effect results from the fact that near the bottom of the potential the
GNMDC term
is not dominant and GR takes over, which  in turn results from the fact that
the more the $\alpha$ parameter
grows
the more dominant becomes the gravitational friction effect and this
allows   the scale
of the theory $\frac{1}{M^{\alpha+1}}$ (needed to produce a long enough
inflation) to decrease significantly. 

Concerning the observables, it can be shown that, for a given value of the scalar power spectrum $\mathcal{P}_{\mathcal{R}}$, while a growing $\alpha$
parameter ameliorates the instability problem, it additionally
affects the values of the spectral index  $n_s$  and the tensor-to-scalar
ratio. In particular, while  $r$ becomes smaller, $n_s$ increases and tends to the outside of the observationally determined Planck likelihood contours, if one
seeks to build a 60 e-fold model \cite{Dalianis:2019vit}.

Similar results can be obtained if one uses as $G(\phi)$ a polynomial 
instead of a monomial form, namely $G(\phi)=\sum_{i}\frac{\alpha_i
\phi^{\alpha_i-1}}{2M_i^{\alpha_i+1}}$. In order for a coupling of such a form 
to produce a different phenomenology than the one studied in the monomial case, 
its various terms must be of comparable magnitude. If this is not the case, then 
the leading monomial term determines the phenomenology.

Finding the scales, $M_i$, so that different terms are comparable is a 
non-trivial task. In \cite{Dalianis:2019vit}, a constraint between the scale 
$M$, the parameter $\alpha$ and the initial values $\phi_*,\dot{\phi}_*$ was 
found (similar to Eq. \eqref{Consinfull} below). This constraint creates a part 
of the phase space that is forbidden, which proves problematic when one seeks 
to build a model with an even value of $\alpha$. This issue is carried over in 
the polynomial GNMDC case, if a term that corresponds to an even value of 
$\alpha$ becomes important, posing yet another problem for the polynomial case. 
However, this is significantly ameliorated in the combined theory proposed in 
this paper (when the NMC term is turned on in Eq. \eqref{Consinfull}), as we 
discuss later.

In summary, if one has a polynomial form of $G(\phi)$ or obtains such a 
polynomial form through quantum corrections \cite{Fumagalli:2020}, the 
non-leading terms will either make an insignificant contribution in 
phenomenology, or in order to affect it  they have to be fine tuned in terms of 
$\alpha$ and $M_i$.

%
%
%
%
%
%

\section{Nonminimal coupling and  generalized nonminimal derivative coupling
combined}
 \label{combined}

 In the previous Section we  presented the inflationary realization of each of
the standalone
modifications to GR, namely of the nonminimal coupling (NMC) and of the  generalized
nonminimal derivative coupling (GNMDC). As we mentioned, the NMC
can lead to observables in very good agreement with observations, however it
possesses the known unitarity problem, while the $\alpha=1$ GNMDC solves the unitarity violation but it leads to  $c_s^2$-instabilities,
while the $\alpha>1$ GNMDC solves the unitarity, but only ameliorates the $c_s^2$ issues while
 making the observable predictions less attractive, in terms of the spectral index.

Keeping the above behaviors in mind, in this section we construct the
combination of the scenarios of  NMC and GNMDC, intending to maintain their separate
advantages while removing their separate disadvantages.

\subsection{The model}

We considered the combined   action   of the form
\begin{align}\label{NMCGNMDCaction}
	S=\int d^4x \sqrt{-g}
\left[\mathcal{L}_{GR}+\mathcal{L}_{\phi}+\mathcal{L}_{NMC}
+\mathcal{L}_{GNMDC}\right],
\end{align}
with
\begin{eqnarray}\label{lagrangianparts}
	&&\mathcal{L}_{GR}=\frac{M_{Pl}^2}{2} R~, \nonumber\\
&&
	\mathcal{L}_{\phi}=-\frac{1}{2} g^{\mu\nu}\partial_\mu \phi
\partial_\nu \phi -V(\phi)~,
\nonumber\\
&& \mathcal{L}_{NMC}=\xi f(\phi) R~,
	\nonumber\\
&&
	\mathcal{L}_{GNMDC}=G(\phi) G^{\mu\nu}\partial_\mu \phi \partial_\nu \phi~.
\end{eqnarray}
Variation in terms of the metric gives rise to the field equations as
\begin{align}\label{fullfeq}
	G_{\mu\nu}&=\frac{1}{M_{Pl}^2}\left[T_{\mu\nu}^{(\phi)}+\xi
T_{\mu\nu}^{(NMC)}-2 G(\phi) T_{\mu\nu}^{(NMDC 1)}\right.\nonumber \\
&\left.\ \ \ \ \ \ \ \ \ \ \ \ -G'(\phi) T_{\mu\nu}^{(NMDC
2)}\right]~,
\end{align}
while variation in terms of the scalar field leads to the   Klein-Gordon
equation
\begin{align}\label{fullKG}
	 \Box\phi&-G_{\mu\nu}\left[ 2 G(\phi) \nabla^\mu \nabla^\nu
\phi+G'(\phi) \nabla^\mu \phi \nabla^\nu \phi \right]\nonumber \\
& +\xi f'(\phi) R-V'(\phi)=0~,
\end{align}
where
\begin{align}\label{fullfeqparts}
	T_{\mu\nu}^{(\phi)}= \nabla_\mu \phi\nabla_\nu
\phi-\frac{1}{2}g_{\mu\nu}\nabla_\lambda \phi \nabla^\lambda \phi-g_{\mu\nu}
V(\phi)~,
\end{align}
\begin{align}
	T_{\mu\nu}^{(NMC)} &=-2 f(\phi)
\left[R_{\mu\nu}-\frac{1}{2}g_{\mu\nu}R\right]\nonumber\\
&-2
f'(\phi)\left[g_{\mu\nu}\Box\phi-\nabla_\mu \nabla_\nu \phi\right]\nonumber\\
&
-2
f''(\phi)\left[g_{\mu\nu}\nabla_\lambda \phi \nabla^\lambda \phi-\nabla_\mu\phi
\nabla_\nu \phi\right]~,
\end{align}
\begin{align}
	&T_{\mu\nu}^{(NMDC 1)} = -G_{\mu\nu}\nabla_\lambda\phi
\nabla^\lambda\phi+4
R^{\lambda}_{\text{ }\text{
}(\mu}\nabla_{\nu)}\phi\nabla_\lambda\phi
\nonumber\\
&
+2
[\nabla^\kappa\phi \nabla^\lambda\phi
R_{\mu\kappa\nu\lambda}
+\nabla_\mu\nabla^\lambda\phi\nabla_\nu\nabla_\lambda\phi-\nabla_\nu\nabla_\mu
\phi\nabla^2\phi]\nonumber \\	
&+g_{\mu\nu}[
\nabla^2\phi\nabla^2\phi-\nabla_\kappa\nabla_\lambda\phi\nabla^\kappa
\nabla^\lambda\phi-2 R_{\kappa\lambda}\nabla^\kappa\phi\nabla^\lambda\phi]\nonumber \\
&-\nabla_\mu\phi\nabla_\nu\phi R~,
\end{align}
\begin{align}
	T_{\mu\nu}^{(NMDC 2)} &= g_{\mu\nu}(\nabla_\lambda\phi\nabla^\lambda\phi
\nabla^2\phi-\nabla^\kappa\phi\nabla^\lambda\phi\nabla_\kappa\nabla_\lambda\phi)
\nonumber\\
&
+2\nabla^\lambda\phi\nabla_{(\mu}\phi\nabla_{\nu)}\nabla_\lambda\phi
-\nabla_\lambda\phi\nabla^\lambda\phi\nabla_\nu\nabla_\mu\phi\nonumber\\
&
-\nabla_\mu\phi
\nabla_\nu\phi\nabla^2\phi~,
\end{align}
with the indices in parentheses denoting
symmetrization. As expected,  for $G(\phi)\rightarrow 0$ we recover the GR+NMC
case, while for
$f(\phi)\rightarrow 0$ we re-obtain the GR+GNMDC case.

Applying the above general field equations in the FRW metric (\ref{FRWmetric})
we extract the    two Friedmann equations as
\begin{align}\label{fried00full}
	\rho_\phi\equiv 3 M_{Pl}^2 H^2&=\frac{\df^2}{2}+V(\phi)+9 G(\phi) H^2
\dot{\phi}^2\nonumber \\
&-6 \xi \left[f(\phi) H^2+f'(\phi) \dot{\phi} H\right],
\end{align}
\begin{align}\label{fried11full}
	-p_\phi &\equiv \MP \left(3 H^2+2
\dot{H}\right)=V(\phi)-\frac{\df^2}{2}+\nonumber \\
&+G(\phi)\left(3H^2
\dot{\phi}^2+2\dot{H}\dot{\phi}^2+4 H \dot{\phi}\ddot{\phi}\right)+2G'(\phi) H
\dot{\phi}^3\nonumber \\
	&-2 \xi \left[ 3 f(\phi) H^2+2 f(\phi)\dot{H}+2 H
f'(\phi)\dot{\phi}+\dot{\phi}^2f''(\phi) \right.\nonumber \\
&\left.+f'(\phi)\ddot{\phi}\right],
\end{align}
where for convenience we have introduced the effective energy density
$\rho_\phi$ and pressure $p_\phi$  for the scalar field.
Additionally, the Klein-Gordon equation (\ref{fullKG}) becomes
\begin{align}\label{KGfull}
	\ddot{\phi} &\left(1+6 G(\phi) H^2 \right)+3 H \dot{\phi} (1+6 G(\phi) H^2+4
G(\phi)\dot{H})\nonumber \\
&+3 H^2 G'(\phi)\dot{\phi}^2-6 \xi f'(\phi)
(\dot{H}+2H^2)+V'(\phi)=0~.
\end{align}
We mention that combining the above equations, one deduces that in order for
the scalar field to obtain real values then the  quantity
\begin{align}\label{Consinfull}
	{\cal{Q}}&\equiv 6 \xi ^2 \df^2 f'(\phi)^2+\xi  \left(2 f(\phi) \df^2+4
f(\phi)
V(\phi)\right)\nonumber \\
&+G(\phi) \left(-3 \df^4-6 \df^2 V(\phi)\right)+\MP \df^2+2 \MP
V(\phi)~,
\end{align}
must be positive.


%
%
%
%
%
%
%
\subsection{Slow Roll Inflation and the three regimes}

From a theoretical perspective, if one investigates  a theory that
combines two different
terms, it is expected that there will be three
different regimes, that one would need to study, depending on their relative
magnitude: one where the GNMDC is
dominating and NMC is a small correction, one that NMC dominates and GNMDC acts
as a small correction, and finally a regime where the two terms are roughly of
the same order. Before discussing   each one individually, and in order to
 facilitate the following  discussion, we  first  provide the general slow-roll
framework of this theory.

In the slow-roll approximation, namely when  $\dot{H} \ll H^2$, $\df \ll H$,
and $\ddot{\phi}\ll 3H\dot{\phi}$, and keeping the leading terms of GNMDC and
NMC,
the first Friedmann equation (\ref{fried00full}) becomes
\begin{align} \label{F00fullSRA}
	3 \MP H^2=9 G(\phi) H^2 \df^2-6 \xi f(\phi) H^2+V(\phi)~,
\end{align}
while  the
Klein-Gordon equation
(\ref{KGfull}) is simplified as
\begin{align} \label{KGfullSRA}
	3H\df \left( 1+6G(\phi) H^2\right)-12H^2 \xi f'(\phi)+V'(\phi)=0~.
\end{align}
Note that regarding $f'(\phi)$ and $f(\phi)$, since we focus in
  monomial  $f(\phi)$ forms which give    $f'(\phi)>f(\phi)$ in
the small field  scenarios ($\phi<M_{Pl}$), we deduce that the difference is
less important than
that between $\df$ and $H$ due to the slow-roll, and hence we  keep only the
$f(\phi)H^2$ term. This approximation will be a posteriori shown to hold in the
numerical analysis of the next section, see Fig. \ref{fig:fulltheorycompare4}.

Using equations \eqref{fried00full} and \eqref{fried11full} we can obtain the
exact form of the  slow-roll parameter
$\epsilon=-\frac{\dot{H}}{H^2}$ as
\begin{align}\label{epsilonanalytic}
\epsilon=\epsilon_{GR}&+\epsilon_{G1}+\epsilon_{G2}+\epsilon_{G3}+\epsilon_{G4}\nonumber \\
&+\epsilon_{N1}+\epsilon_{N2}+\epsilon_{N3}+\epsilon_{N4}~,
\end{align}
where we have introduced the following auxiliary parameters
\begin{align*}
 \epsilon_{GR}  = \frac{\df^2}{2 \MP H^2}~,
\end{align*}
\begin{eqnarray}
 &&   \epsilon_{G1}
= \frac{3
\df^2 G(\phi)}{\MP}~,   \quad   \epsilon_{G2}  = -\frac{\df^2 \dot{H}
G(\phi)}{\MP
H^2}~,  \nonumber \\
&&
 \epsilon_{G3}  = -\frac{2\df \ddot{\phi} G(\phi)}{\MP H}~,     \quad
\epsilon_{G4}  = -\frac{G'(\phi)\df^3}{\MP H}~,
 \nonumber \\
&&
\epsilon_{N1}  =
\frac{2\xi f(\phi) \dot{H}}{\MP H^2}~,   \quad
 \epsilon_{N2}  = -\frac{\xi f'(\phi) \df}{\MP H}~,
 \nonumber\\
 &&
 \epsilon_{N3}  =
\frac{\df^2 \xi f''(\phi)}{\MP H^2}~,    \quad   \epsilon_{N4}  = \frac{\xi
f'(\phi) \ddot{\phi}}{\MP H^2}~.
\label{auxilepsilons}
\end{eqnarray}
These separate parameters will be useful in order to quantify which specific
term of the theory dominates the inflationary realization, and in particular
the $\epsilon_{Gi}$   are related to the GNMDC while the
$\epsilon_{Ni}$
  are related to the NMC ($i$   runs from 1 to 4), while $\epsilon_{GR} $ is
the usual slow-roll parameter of  minimally coupled, single-field inflation.
  From  our previous discussion it becomes clear that in the slow-roll
era the only important terms should be $\epsilon_{G1}$, $\epsilon_{N1}$ and
$\epsilon_{N2}$.

We can now move on to calculate the various perturbative functions,
  as functions of the auxiliary parameters defined above. Using the definitions in
Appendix \ref{perturbations}  we find
\begin{align}
&\mathcal{G}_T=\MP\left(1-\frac{\epsilon_{G1}}{3}-\frac{\epsilon_{N1}}{\epsilon}
\right)~.
\\
&\mathcal{F}_T=\MP\left(1+\frac{\epsilon_{G1}}{3}-\frac{\epsilon_{N1}}{\epsilon}
\right)~,
\\
&	\Sigma=\MP
H^2\left(\epsilon_{GR}+6\epsilon_{G1}+6\epsilon_{N2}+3\frac{\epsilon_{N1}}{
\epsilon }
-3\right)~,
\\
&	\Theta=\MP H
(1-\epsilon_{G1}-\epsilon_{N2}-\frac{\epsilon_{N1}}{\epsilon})~,
\end{align}
\begin{eqnarray}
&&	
\!\!\!\!\!\!\!\!\!\!\!\!\!
\mathcal{G}_s=-\frac{\MP}{9}
		\left[\eniena+\epsilon(\egena+\enidio-1)\right]^{-2}
		 \nonumber \\
	&&\!\!\!\!\cdot   \left[ \epsilon
(\egena-3)+3\eniena\right]\left\{-3\eniena
(\egena+\egr)
 \right. \nonumber \\
&& \! \!\!\!\left. \! +
\epsilon\! \left[
3\egena^2\!+\!3\egr\!+\!9\enidio^2
+\egena (3\!-\!\egr\!+\!12 \enidio)\right]\!\right\}\! ,
\end{eqnarray}

\begin{eqnarray}
	&&
	\!\!\!
	\mathcal{F}_s =
	-\frac{\MP}{9 [\epsilon(\egena+\enidio-1)+\eniena]^2}
\notag \\
	&&
	\ \ \
	\cdot
	\Big\{ \epsilon^2 \left\{\egena^2 \left[7
\eniena+17
\enidio+\enitria-4\right.\right.
\notag \\
	&&\left.	\ \ \ 	\ \ \ 	\ \	\ \ \ 	\ \ 	\ \ \
	-3\left(\egtria+\egtessera+\enitessera\right)-4\right]
  \nonumber \\
&&  	\ \
	\ \ \ +4 \egena^3+3 \egena \left[2 \egtria-2 \egtessera
(\enidio-1)-10 \eniena\right.
\notag    \\
	&& \left.
	\ \ \ \,\ -2 \enitria
	+5 \enidio (\enidio-2) +2 \enitessera\right]+9
\left(2 \egtessera \enidio  \right.\nonumber \\
	&&
	\ \ \
	\ \left. \left.+\egtria+\egtessera+3 \eniena-3
\enidio^2+\enidio+\enitria+\enitessera\right)\right\}
\notag
 \\
	&&
	\ \ \ \
	\left.+\epsilon \eniena \left[\egena (-6 \egtria-6 \egtessera+15
\eniena+30
\enidio+6 \enitria \right.\right.\nonumber \\
&&\left.\left.
\ \ \ \ \ \ \ \ \,\ \ \ \
-6 \enitessera)-9 \left(2 \egtria+2 \egtessera
(\enidio+1)+3 \eniena \notag \right.\right.\right.\\
	&&
	\ \ \ \ \ \ \ \ \,\ \ \ \
	\left.\left.+2
(\enidio+\enitria+\enitessera)-3
\enidio^2\right)+4 \egena^2\right]
\nonumber \\
& & \ \ \ \
+\epsilon^3 (\egena-3)^2
(\egena-1) \nonumber
\\
& & \ \ \ \
+9 \eniena^2
(\epsilon-\epsilon_{GR}-\egena-\egdio)\Big\}.
\end{eqnarray}

We can now proceed to calculate the soundspeed of the theory. If we insert these
equations into
the definition of the sounspeed  \eqref{sspddef}  we obtain
\begin{eqnarray}
&&
\!\!\!\!\!\!
c_s^2 =
\left\{\epsilon
\left[\egena
(12 \enidio-\epsilon_{GR}+3)+3 \left(\egena^2+\epsilon_{GR}+3
\enidio^2\right)\right]\right.
\notag \\
&& \ \ \ \  \,
\left.-3
\eniena (\egena+\epsilon_{GR})\right\}^{-1}
\left[\epsilon (\egena-3)+3 \eniena\right]^{-1}
\notag \\
&&
 \ \ \    \,
 \cdot \Big\{\epsilon^2 \left\{\egena^2 \left[7 \eniena+17
\enidio+\enitria\right.\right.
 \notag \\
&&
 \  \ \ \ \  \, \ \ \  \ \ \
 \left.
-3 \left(\egtria-\egtessera-\enitessera\right)-4\right]
\notag \\
&&  \ \ \  \ \  \
+3 \egena \left[5 (\enidio-2)\enidio -2 \egtessera
(\enidio-1)   -10 \eniena\right.
\notag \\
&& \ \ \ \  \, \ \ \ \ \ \ \ \ \ \  \, \left.
+2
\left(\egtria-\enitria+\enitessera\right)\right]
 \notag \\
&&  \ \  \ \ \    \
+4 \egena^3+9 \left(\egtria+2
\egtessera
\enidio+\egtessera+3 \eniena\right.
 \notag \\
&& \left.\left.\ \  \ \ \
-3
\enidio^2+\enidio+\enitria+\enitessera\right)\right\} \notag \\
&&
\ \  \ \ \ \,
 +\epsilon \eniena \left\{\egena \left[6\left(
\enitria- \egtria- \egtessera- \enitessera\right)\right.\right.
\notag \\
&& \left.\ \  \ \ \  \ \ \ \,  \ \  \ \ \ \,
+15 \eniena+30
\enidio\right]
\notag \\
&&  \ \  \ \ \ \,
+4 \egena^2-9 \left[2 \egtria+2 \egtessera (\enidio+1)+3
\eniena \right.  \notag \\
&&\left.\left.
\ \  \ \ \  \ \ \ \,  \ \  \ \ \
+2 (\enidio+\enitria+\enitessera)-3
\enidio^2\right]\right\}
\notag \\
&& \ \  \ \ \ \,
+\epsilon^3 (\egena-3)^2 (\egena-1)
\notag \\
&&  \ \ \    \,
+9 \eniena^2
(\egtria+\egtessera+\eniena+\enidio+\enitria+\enitessera)\Big\},
\label{soundspeedgene}
\end{eqnarray}
which is an exact expression. Let us consider its various limits. First,
in the   GR
limit,
where $\epsilon_{Gi}$, $\epsilon_{Ni} \rightarrow 0$, we can see that $c_s^2 $
becomes
identically equal
to 1 as expected. The same holds in the NMC limit, where $\epsilon_{Gi}
\rightarrow 0$, again as expected. Moreover, in the
GNMDC limit, where $\epsilon_{Ni}\rightarrow 0$, we acquire
\begin{align}\label{csgnmdclimit}
	{c_s^2}&=\frac{1}{(\egena-3) (3 \egena (\egena+1)-(\egena-3)
\epsilon_{GR})}\cdot \nonumber \\
&\cdot \left\{ \epsilon (\egena-1) (\egena-3)^2+9
(\egtria+\egtessera)+\right.\nonumber \\
&\left.\egena \left[\egena (4
\egena-3 \egtria-3 \egtessera-4)+6 (\egtria+\egtessera)\right]\right\},
\end{align}
which using the definitions (\ref{auxilepsilons}) gives expression
\eqref{cssqGRNMDC}.

However, in general  we would  like to extract more information about the
behavior of the full equation \eqref{soundspeedgene}. A detailed manipulation of
this equation is quite tedious and is not included here, however there is a
clear note to be made based on it. If we use  equation \eqref{epsilonanalytic},
to substitute $\epsilon$ with the auxiliary $\epsilon$ functions, we  end up
with an expression of the form
\begin{align}\label{csorder}
	c_s^2-1\approx\frac{\mathcal{O}(\epsilon_{Gi})}{f_{\epsilon}(\epsilon_{Ni},
\epsilon_{GR})+\mathcal{O}(\epsilon_{Gi})},
\end{align}
where    $f_{\epsilon}$ is a function that does not depend on the
$\epsilon_{Gi}$ while $\mathcal{O}(\epsilon_{Gi})$ is a function at least linear in $\epsilon_{Gi}$. Hence, the denominator is of greater order of magnitude as compared to the nominator of this fraction, when Slow Roll has ended and the NMC terms completely take over, if one chooses a derivative
coupling that vanishes towards the bottom of the potential. This implies that
$c_s^2 \equiv 1$, which is, in fact, one of the main results of this work: the
inclusion of NMC and a vanishing $\phi$-dependent GNMDC, regardless   
its exact form, can completely heal the $c_s^2$ instabilities
of derivative coupling (see also Fig. \ref{fig:fulltheorycompare3} for
corresponding numerical results). This was expected because
the NMC term has a sound speed identically equal to 1 and its terms remain
 dominant after the end of the slow roll.

On the other hand,
using the same rationale with equation \eqref{csgnmdclimit}, we can show that
in the GNMDC limit  we acquire
\begin{align}
	c_s^2-1\approx\frac{\mathcal{O}(\epsilon_{Gi})}{\mathcal{O}(\epsilon_{Gi})}.
\end{align}
This fraction, is obviously non-zero in general, and in particular it can be
larger or smaller than 1. This  reconfirms the results
of \cite{Dalianis:2019vit}, regarding the squared sound-speed oscillations between
negative and superluminal values.

%
%
%
%
%
%
\subsubsection*{Regime 1: \textbf{NMC $\gg$ GNMDC}}

If one would like to study the case where the GNMDC term is negligible
compared to the NMC term during the slow-roll era, then observing  equations
\eqref{F00fullSRA}, \eqref{KGfullSRA},  there are two requirements that should
be satisfied, namely
\begin{align}
	\xi \gg \frac{G(\phi)}{f'(\phi)}H\df~,
\end{align}
and
\begin{align}
	\xi \gg \frac{G(\phi)}{f(\phi)}\df^2~,
\end{align}
where, based on our previous
discussion, we deduce that the former is actually stronger than the latter.
Nevertheless, we should mention here that the GNMDC form (\ref{Fphiform})  on
which
we will
focus on in this work,  turns off at the end of inflation, hence,  if we
enforce the above  constraints then the GNMDC will be unimportant throughout
the field's evolution. Thus, this case   would bear
practically no effect in both the early and late stages phenomenology, and
therefore we will not investigate it further.
%
%
%
%
	
%
%
%
\subsubsection*{Regime 2: \textbf{GNMDC $\gg$ NMC}}

In order to realize this regime of  GNMDC domination, using equations
\eqref{F00fullSRA}   and \eqref{KGfullSRA}
  we extract the requirements
\begin{align}
	\xi \ll \frac{G(\phi)}{f'(\phi)}H\df~,
\end{align}
and
\begin{align}
	\xi \ll \frac{G(\phi)}{f(\phi)}\df^2~,
\end{align}
where the latter is stronger than the former if the dynamics of the NMC are to
be negligible in the slow-roll  era. However, unlike the previous case where
GNMDC $\ll$ NMC,
the post-slow-roll dynamics cannot be studied without the NMC terms. This is
due to
the fact that a $\phi$-dependent GNMDC quickly becomes subdominant near the bottom of the
potential, in the post-slow-roll regime. Hence, this case should be studied in
more detail, and in particular to examine
 the
sound
speed squared, since in the sole GNMDC model the
derivative coupling has been shown to create instabilities
due to $c_s^2<0$. 

Specifically, as discussed and shown with eq. \eqref{csorder}, we are interested in
investigating   whether the inclusion of the NMC
term corrects the $c_s^2$ values towards 1,   compared to the standalone
 GNMDC case.
 A theoretical indication towards this direction is
  that the NMC sound speed is identically
equal to 1, and since the NMC should take over (or at least be comparable) with
GNMDC in the post-slow-roll period  it is expected that the sound speed will be
corrected, albeit the magnitude of this correction remains to be found, since eq. \eqref{csorder} is only qualitative.
Instead,
of providing explicit results here, we will do it in the analysis of the next
regime, namely where NMC $\approx$ GNMDC.

\subsubsection*{Regime 3: \textbf{NMC $\approx$ GNMDC}}

We can now proceed to the investigation of the case where NMC and GNMDC terms
are of the same order. We start with the slow-roll dynamical equations
presented above. To enforce NMC
$\approx$ GNMDC  we can choose between the two requirements presented earlier,
one of which is stronger. For simplicity we will choose the weaker constraint
which nevertheless is adequate  for the results of our model. In particular,
we
enforce
\begin{align}
	\xi f'(\phi)\approx G(\phi) H \df~,
\end{align}
while still
\begin{align}
	\xi f(\phi) \gg G(\phi) \df^2~,
\end{align}
and we additionally require that the GR terms are negligible   compared to the
GNMDC and
NMC ones during the slow-roll era. Then, the scalar-field equation
\eqref{KGfullSRA} becomes
\begin{align} \label{KGfullSRAnext}
	18 G(\phi) H^3  \df+V'(\phi)=12 H^2 \xi f'(\phi)~,
\end{align}
while the Friedmann equation \eqref{F00fullSRA} is significantly simplified and
becomes
\begin{align} \label{F00fullSRAnext}
	3 \MP H^2+6 \xi f(\phi) H^2=V(\phi)~.
\end{align}

Based on the discussion following equations \eqref{F00fullSRA}   and
\eqref{KGfullSRA},
regarding the slow-roll approximations,   the dominant parameters
during the slow-roll  period are $\egena$, $\eniena$ and $\enidio$. Hence, if
we are
interested in the early phase's predictions we can, as a first approximation,
keep only the first-order contributions with regards to these parameters. We
thus acquire
\begin{align}
	\mathcal{F}_s=\mathcal{G}_s \approx \MP \egena~.
\end{align}
Inserting this into the squared sound-speed relation
(\ref{sspddef})   we obtain $c_s^2=1$ during the slow-roll
period (equivalently maintaing only  $\egena$, $\eniena$ and $\enidio$   in the
general expression (\ref{soundspeedgene}) gives $c_s^2=1$).

We proceed by  calculating the inflationary observables. Using expression
(\ref{PsandPtdef}), for the power spectrum at
first order we obtain
\begin{align}
	\mathcal{P}_\mathcal{R} \approx \frac{H^2}{8\MP\pi^2 \egena}~,
\end{align}
which coincides with   \eqref{PsinGRNMDC} if one keeps only the first
order contribution. Interestingly enough, at first order the NMC term does not
have an effect on the scalar power spectrum value, since the only $\epsilon$
parameter appearing is $\egena$.

Concerning  the tensor-to-scalar ratio  $r$, using
(\ref{PsandPtdef}), (\ref{rdef}) we find
\begin{align} \label{rprediction}
	r &\approx  16 \egena+\frac{16\egena}{\egena+\enidio}\eniena~.
\end{align}
Unlike the scalar power spectrum, this result clearly shows the effect of the
combined theory. In particular, during slow-roll we have $\eniena<0$, which
implies that  the NMC term lowers the
standard result, which is $r=16 \egena$, namely improving the tensor-to-scalar
ratio to values that are in better agreement with the observations. Hence, the
very low tensor-to-scalar ratio, which is a characteristic result
of the NMC term, is maintained in the combined theory.

Concerning the scalar spectral index, $n_s$, using
(\ref{PsandPtdef}), (\ref{nsdefgene}) in the case of the present combined
scenario we obtain:
\begin{align}
	n_s \approx 1+\frac{-3
(\egtria+\egtessera)+2\egena(\egena+\eniena+\enidio)}{
\egena(\egena+\eniena+\enidio-1)}~.
\end{align}
Note that here  we cannot ignore the terms $\egtria$, $\egtessera$ as we have
done until now, since the rest of the terms are of second order in the
$\epsilon$ parameters.  As expected, when the NMC-related parameters go to zero
we can recover the result  (\ref{nsinGRNMDC}) of the standalone GNMDC model.

In summary, when a $\phi$-dependent GNMDC and the NMC terms are
comparable, the scenario   can be completely healed from the
$c_s^2<0$ unstable region. Additionally, the value of the tensor-to-scalar
ratio
not only remains inside the Planck 2018's contour plots, but it is
 increasingly improving  as the NMC contribution
becomes more significant. Finally, the scenario can, in principle, be healed from the unitarity
problem, because as the GNMDC term becomes more significant in the   
slow-roll period, the magnitude of $\xi \phi_*^2$ decreases significantly. 
These features make the combined scenario at hand better than its
individual counterparts. We emphasize that all   the above results hold  as 
long as   $G(\phi)$ is $\phi$-dependent, and thus at the bottom of the 
potential it becomes negligible.

%
%
%
%
%
%
\section{Numerical investigation}
\label{specificstandalones}

In this Section we perform a full numerical study, in order to 
demonstrate, by use of specific examples, the general results of our theory 
obtained in the previous Section, most importantly equations \eqref{csorder} and 
\eqref{rprediction}. To satisfy the ansatz that GNMDC becomes negligible at the 
end of 
inflation we choose a monomial or polynomial form for $G(\phi)$, however  other 
similar forms  still produce viable results.

To numerically elaborate, we consider, then, 
specific NMC and GNMDC functions. For the former, i.e for the coupling function
$f(\phi)$ we choose the most well-studied case of the standalone NMC  scenario,
namely  $f(\phi)=\xi \phi^2$, while for the latter we consider the well-studied monomial
form (\ref{Fphiform}), namely $
	G(\phi)=\frac{\alpha
\phi^{\alpha-1}}{2M^{\alpha+1}}$~. We then provide and discuss an example with a polynomial form $
	G(\phi)=\sum_{i}\frac{\alpha_i
\phi^{\alpha_i-1}}{2M_i^{\alpha_i+1}}$.
Additionally, in order
to have increased theoretical justification, and to compare with the
literature, we consider the scalar field to be the   Higgs
boson and thus its potential to be the known quartic Higgs one
\cite{Germani:2010gm,Atkins:2010yg},
namely
\begin{align}
	 V(\phi)= \frac{\lambda \phi^4}{4}.
\end{align}
Finally, in what follows, we have imposed the normalization
 that the scalar power spectrum value at $k=0.05
Mpc^{-1}$ is $\mathcal{P}_\mathcal{R}=2.2\cdot10^{-9}$ \cite{Akrami:2018odb}.
Additionally, the initial conditions are
selected in order for the produced models to yield 40, 50 and 60 e-folds.

\begin{figure}[ht]
\center
\includegraphics[width=\linewidth]{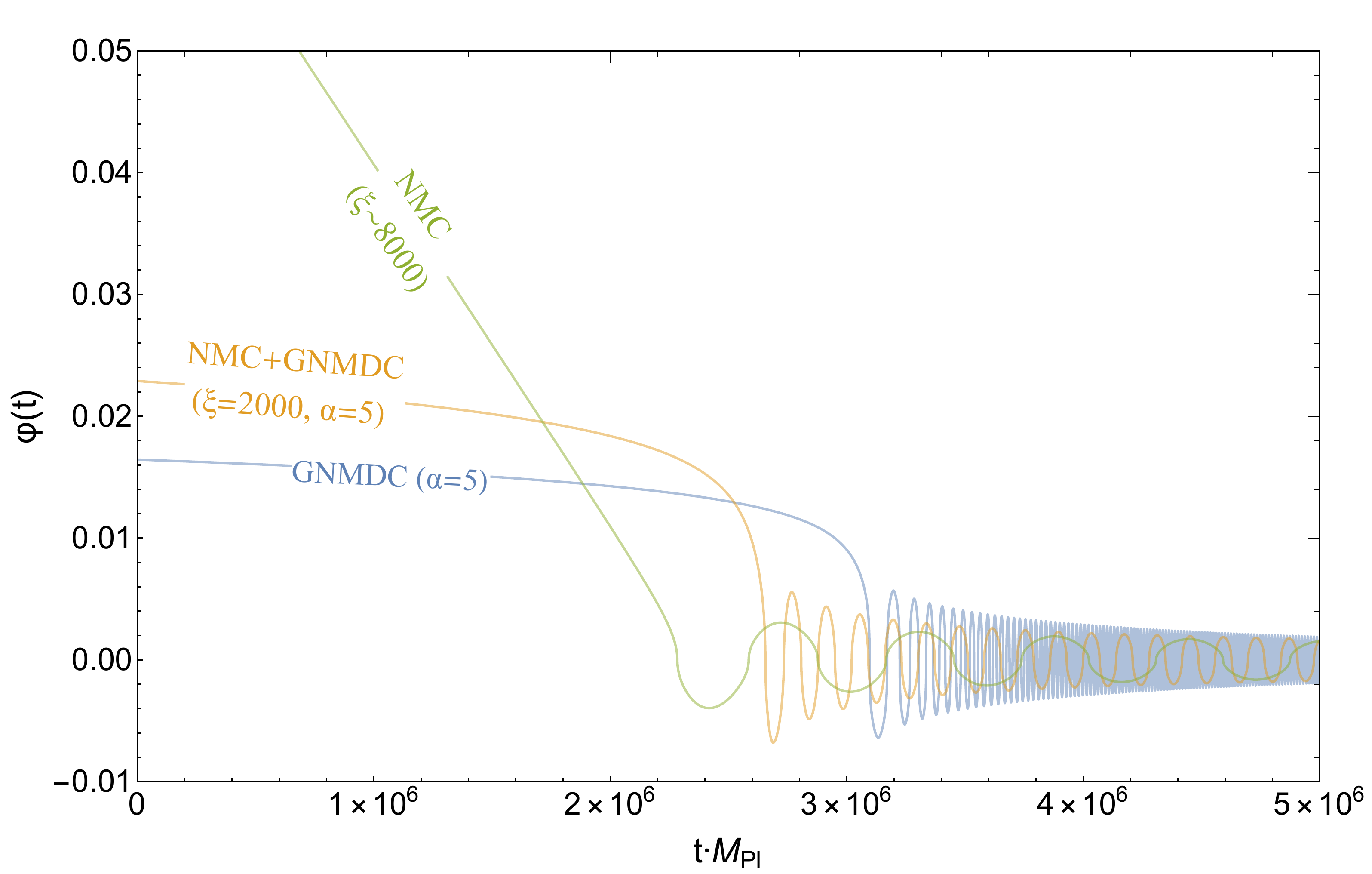}%
\\[0.4cm]
\caption{{\it{The evolution of the scalar field in three different cases: For
 GNMDC with $\alpha=5$, for NMC with $\xi\approx 8000$ and for the
combination of NMC+GNMDC with $\alpha=5$ and $\xi=2000$, respectively. All
three models yield 60 e-folds and $\mathcal{P}_\mathcal{R}=2.2\cdot10^{-9}$. One can
observe the lengthening of the period of oscillations, as well as the larger
initial value of the field, in the case where NMC becomes more important ($\xi$
grows). Finally, note that when
the two theories are combined, $\xi \phi_*^2$ remains less than $M_{Pl}$.}}}
\label{fig:fulltheorycompare1}
\end{figure}

Starting with the monomial GNMDC scenario, in Fig. \ref{fig:fulltheorycompare1} we depict the evolution of the scalar
field   for the various   cases.
 The main observation from this graph
is the fact that although in the standalone GNMDC scenario the
oscillations of $\phi$ (and consequently of $\dot{\phi}$) are quite wild, in
the combined scenario  the period of the field oscillations increases. This
will play a crucial role in the
following analysis since it is the cause of the $c_s^2$-instabilities healing
in the combined scenario.
\begin{figure}[!]
\center
\includegraphics[width=\linewidth]{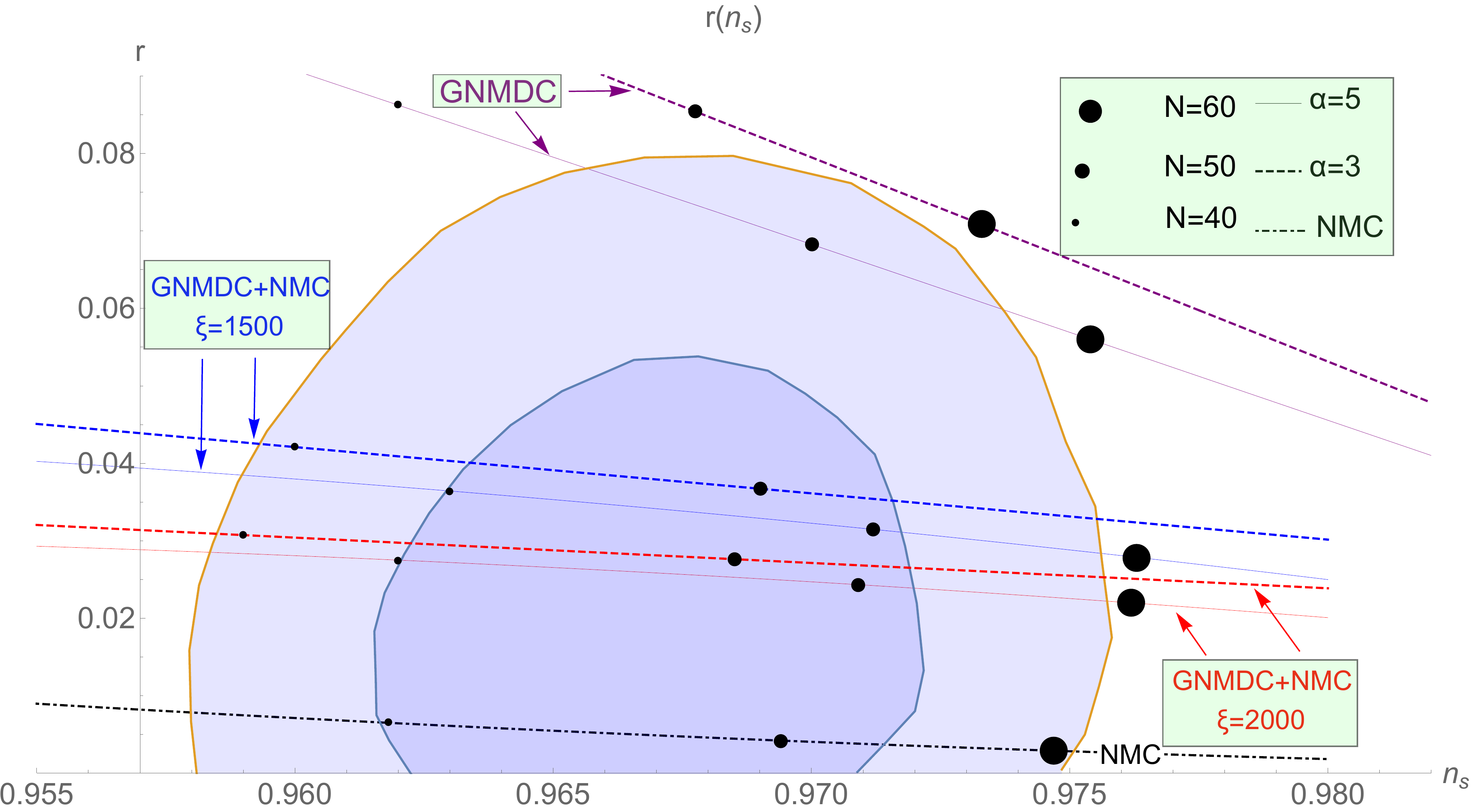}
\\[0.4cm]
\caption{{\it{
 1$\sigma$ (purple) and 2$\sigma$ (light purple) contours for Planck 2018
results (Planck $+TT+lowP$)  \cite{Akrami:2018odb}, on the
$r-n_s$ plane, alongside the predictions of the scenarios at hand.
  The NMC
scenario
corresponds to the dot-dashed line. For the  GNMDC (purple lines) we have
chosen to show two cases, one with $\alpha=3$ (dashed line) and one with
$\alpha=5$ (dotted line). The same convention is used for the combined
 NMC+GNMDC scenario in terms of $\alpha$, where we also have the colour code
of blue lines for $\xi=1500$ and red lines for $\xi=2000$. It is evident that
the
NMC term lowers the $r$ value as it becomes more significant, as compared to
the
GNMDC alone. Very low $r$-values are a main feature of NMC of the form
$\phi^2$.
Moreover, one observes that for the same value of $\xi$, as $\alpha$ grows  $r$
is also lowered, as   reported in \cite{Dalianis:2019vit} too. Finally, the
growing dots
represent 40, 50 and 60 e-folds respectively.}}}
\label{fig:fulltheorycompare2}
\end{figure}

As a next step we calculate the inflationary
 observables, and in particular the scalar spectral  index and the tensor to
scalar ratio, using the exact expressions of Appendix \ref{perturbations}.
In Fig. \ref{fig:fulltheorycompare2}, we present the obtained results for the
standalone cases of NMC and of GNMDC, as well as for the combined scenario.
Additionally, for transparency, in the same figure we provide the 1$\sigma$ and
2$\sigma$ contours of the Planck 2018 data \cite{Akrami:2018odb}.
As we observe, the simple NMC gives very satisfactory predictions however due
to the unitarity violation this model has to be abandoned. The simple GNMDC
scenario solves the unitarity issue however it leads to quite large $r$ values
and moreover it leads to instabilities related to $c_s^2$. We
observe that, in the combined NMC+GNMDC scenario which alleviates the unitarity
issue, one can improve the obtained $r$ values, bringing them back inside the
Planck 2018 contours, and moreover the larger the $\alpha$ value is the larger
is the improvement.   Specifically, one observes that
for the same value of $\alpha$ (dashed lines for $\alpha=3$, dotted lines
for $\alpha=5$), as $\xi$ grows the tensor-to-scalar ratio lowers. Moreover,
  for the same value of $\xi$ (blue lines for $\xi=1500$, red lines for
$\xi=2000$), as $\alpha$ grows, $r$ also lowers. This result is also expected
since this is one of the effects of the sole GNMDC term
\cite{Dalianis:2019vit}.

In conclusion, monomial GNMDC models with larger
$\alpha$ would in fact be more desirable in the context of the combined theory
proposed in this work, due to the enhancement of the gravitational friction
effect that the $\alpha$ parameter essentially quantifies. However, if one considers a polynomial GNMDC
the same effect can actually be obtained, since inflation can be carried by two 
or more  ``frictious'' terms present in a polynomial GNMDC. We demonstrate such 
a scenario later.

\begin{figure}[ht]
\includegraphics[width=\linewidth]{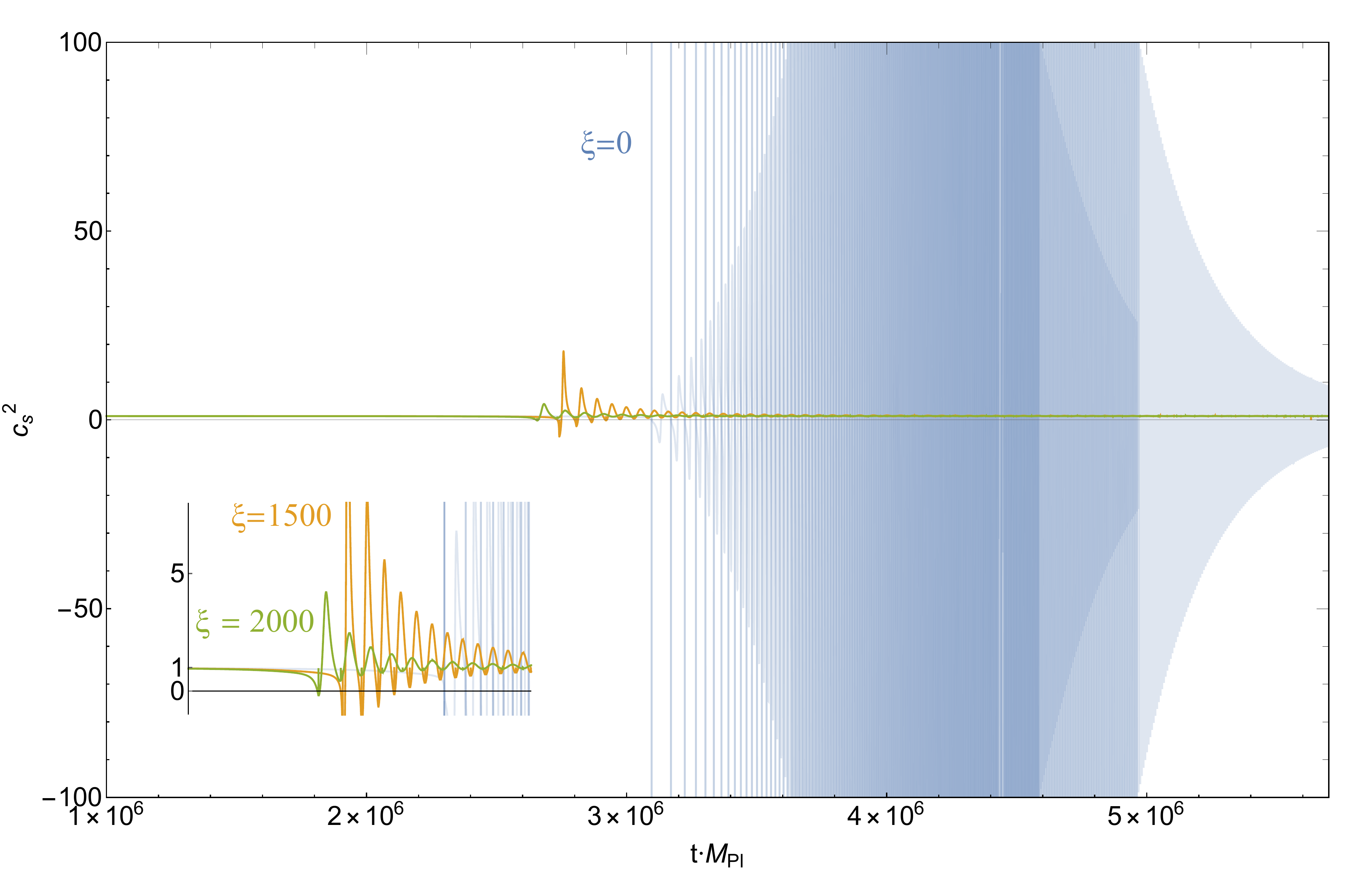}%
\\[0.4cm]
\caption{{\it{The squared sound speed evolution for the scenario at hand,
for  $\alpha=5$, with $\xi=0, 1500$ and $2000$ respectively. It is clear that
as
the NMC contribution becomes more significant ($\xi$ increases), the
oscillations in its value are damped and $c_s^2$ is corrected towards 1.}}}
\label{fig:fulltheorycompare3}
\end{figure}

\begin{figure}[!]
\includegraphics[width=\linewidth]{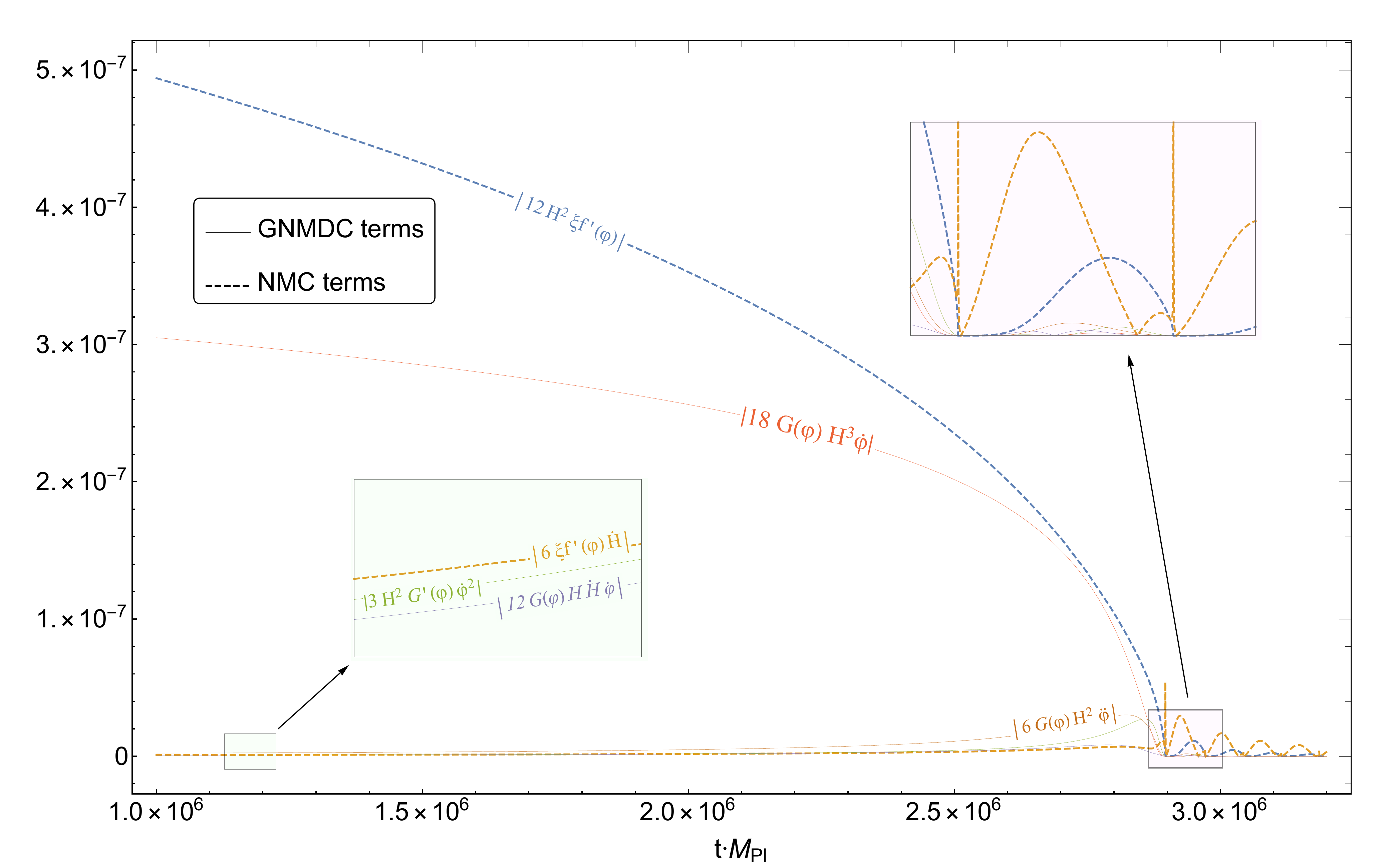}%
\\[0.4cm]
\caption{{\it{
The
contribution of the terms related to the GNMDC (continuous
lines) and to the NMC (dashed lines) in the
Klein-Gordon    equation (\ref{KGfull}), for $\alpha=5$  and
$\xi=2000$.
It is clear than when the oscillations start, the NMC terms dominate over the
GNMDC terms and hence the GNMDC effects are negligible, even though during the
slow-roll era they are comparable. The standard GR terms are intentionally
omitted in order to
make the graph simpler.}}}
\label{fig:fulltheorycompare4}
\end{figure}

Let us now examine the   evolution of $c_s^2$ in order to verify that the
combined scenario can indeed heal the $c_s^2$-instabilities of the standalone
GNMDC.
In Fig. \ref{fig:fulltheorycompare3}  we depict the evolution of $c_s^2$ for
various cases. As one can clearly see, while in the standalone GNMDC (i.e. for
$\xi=0$) the  $c_s^2$ wildly oscillates between positive and negative values,
when we switch on the NMC contribution we obtain a significant decrease of the
oscillatory behavior and a stabilization to positive values.
In particular, in the combined scenario we observe that near the end of
inflation the GNMDC contribution smooths out, while the NMC
term remains co-leading alongside the standard GR (i.e. of the minimally
coupled scalar field) terms.
However, it is known that the standalone NMC as well as the  GR
terms have no instability issues. Hence, the $c_s^2<0$ problem is healed.

In order to provide a more transparent picture of the above relative effect of
the GNMDC and the NMC contributions, in Fig. \ref{fig:fulltheorycompare4} we
present the
contribution of the terms related to the GNMDC and the NMC in the
Klein-Gordon    equation (\ref{KGfull}). From this graph it becomes clear that
although during the slow-roll era the contributions from NMC and GNMDC are
comparable,  when the oscillations start the NMC terms dominate completely
over the GNMDC terms and since NMC alone leads to   $c_s^2=1$ its complete
dominance in the combined model is adequate to bring $c_s^2$ away from the
unstable region (caused by standalone GNMDC).  Hence,
the GNMDC contribution to the $c_s^2$ at the end of inflation is overpowered by
the NMC contribution and the wild oscillations of the sound speed are damped
much earlier in this scenario. Note that this damping of oscillations is
more efficient for larger $\alpha$ values, which as we mentioned above lead
also to better $r$ values. Hence, overall, larger $\alpha$ values would be more
desirable. 

This brings up the question of whether this is a realistic scenario. 
Quantum corrections should, in fact, bring about terms that might be of lower 
order, so one should check the resulting phenomenology. However, if one chooses 
a polynomial form for $G(\phi)$, a very similar phenomenology occurs, since, 
unless the various terms of the polynomial are finely tuned, there will still be 
only one monomial term that drives the slow roll and hence it will produce very 
similar results.

Moreover, even in the case that two, or more, terms are actually of the 
same order of magnitude, the phenomenology is still, qualitatively, the same, 
since the results of Section \ref{combined} are independent of the exact form of 
the GNMDC. They are only based on the fact that the GNMDC should  become 
negligible at the end of inflation. Nevertheless, we later provide a numerical 
example of such a scenario for demonstrative purposes.

In conclusion, in the combined scenario, when inflation starts, the
gravitational friction effect due to
the GNMDC term is what causes the model to produce a significant amount of
e-folds without having to resort to $\xi \phi_*^2>M_{Pl}$ values as in the
standalone NMC case, and this is what alleviates the unitarity issue. At the
same time, the NMC term causes the $r$ of the model
to be significantly lowered, and thus be in better agreement with observations, as
compared to the standalone GNMDC case. Finally,
when inflation ends and the oscillations start, the NMC terms remain more
significant than that of the GNMDC, which leads to the fast eradication of the
oscillations in the $c_s^2$ value, healing the theory of instabilities. These
features  and advantages of the combined scenario are amongst the main results
 of the present work.

Before closing this section, we note another role of the GNMDC parameter
$\alpha$ on the results. In the combined scenario even
$\alpha$  values  can still lead to viable
inflation. This  is not the case when GNMDC is considered alone
\cite{Dalianis:2019vit} due to its inability to satisfy the corresponding
requirement
(\ref{Consinfull}), which essentially disqualifies the area of the phase
space corresponding to desirable observables. The fact that in the combined
scenario all $\alpha$ values can be used,  is a significant advance in the
richness of the resulting phenomenology. 

This also holds for a polynomial GNMDC form: 
when an even-valued $\alpha$ term becomes important the polynomial GNMDC numerics become
unstable due to the above constraint. This can be ameliorated or even healed, when it is combined with NMC.
%
%
%
%
%
%
%

As a further numerical demonstration of the effects of the NMC+GNMDC scenario we include a model resulting from a polynomial GNMDC form, namely
\[
	G(\phi)=\sum_{i}\frac{\alpha_i
\phi^{\alpha_i-1}}{2M_i^{\alpha_i+1}},
\]
where $i$ is a subscript that defines which and how many corresponding terms are taken into account. As a demonstration we pick
\begin{align} \label{polynomialGNMDC}
	\frac{\alpha  \phi ^{\alpha -1}}{2 M_1^{\alpha +1}}+\frac{(\alpha -1) \phi 
^{\alpha -2}}{2 M_2^{\alpha }},
\end{align}
with $\alpha=4$, while the NMC coupling is $\kappa \approx 2000$. The coupling coefficients should be such that these two terms as a whole are of comparable magnitude. If not, then one of the terms would dominate during slow roll, essentially reducing the model to the monomial form presented earlier. We reiterate, that such a scenario is not viable in the sole GNMDC case, since even values of $\alpha$ are problematic.

A polynomial form still falls within the ansatz needed for the results 
of Section \ref{combined} to hold, namely that the GNMDC becomes negligible near 
the end of inflation. The results shown therein, then, still hold, since this 
modification affects only the exact form of the $\epsilon_{G_i}$ parameters, and 
not their overall behavior.

\begin{figure}[!]
\center
\includegraphics[width=\linewidth]{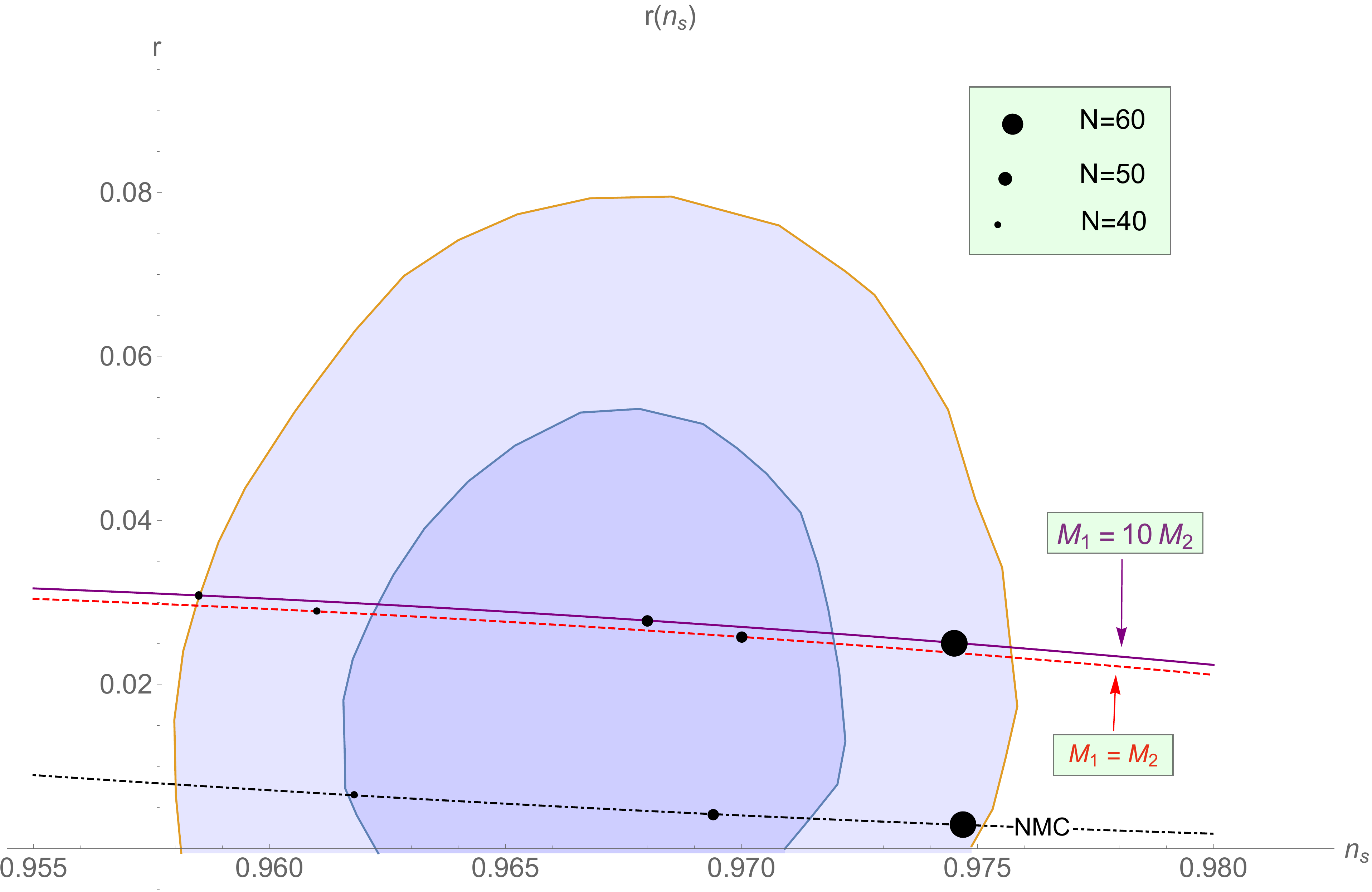}
\\[0.4cm]
\caption{{\it{
 1$\sigma$ (purple) and 2$\sigma$ (light purple) contours for Planck 2018
results (Planck $+TT+lowP$)  \cite{Akrami:2018odb}, on the
$r-n_s$ plane, alongside the predictions of the polynomial GNMDC+NMC scenario.
To demonstrate that the resulting phenomenology is very similar to the 
monomial case of Fig. \ref{fig:fulltheorycompare2}, we present two cases, one 
with $M_1=M_2$ and one with $M_1=10M_2$. The
growing dots
represent 40, 50 and 60 e-folds respectively. The corresponding models when NMC 
is not included are not stable  due to the constraint \eqref{Consinfull}.}}}
\label{fig:polynomialrns}
\end{figure}

Definitely, picking initial conditions and scales for the combined 
theory, with a polynomial GNMDC, is a tedious task as compared to the monomial 
case. Nevertheless, by imposing the ansatz discussed earlier, regarding the 
magnitude of the various terms of the GNMDC, one can obtain results that are 
well within observational bounds. The overall picture is very similar to the 
monomial case, as one can  see in Fig. \ref{fig:polynomialrns}, which  
was expected since the $\epsilon_{G_i}$ show a similar behavior.

\section{Conclusions}\label{conclusions}

It is widely accepted that for modern Cosmology to explain the hot big bang and
the primordial perturbations observed through the CMBR it has to be complemented
by an initial inflationary period. There is a variety of ways to achieve
inflation,
the most well-studied of which is the inclusion of a scalar field, that through
its dynamics affects the evolution of the infant Universe. Given that the only
scalar field actually observed in nature up to now is the Higgs field, it would
be the prime candidate for such a scenario.

The basic scenarios in which the Higgs field is minimally coupled to gravity
have been excluded from Planck observations \cite{Akrami:2018odb}. The next
candidate is to allow the Higgs field to couple nonminimally (NMC) with gravity.
NMC Higgs inflation, with a quadratic coupling of the form $f(\phi)=\phi^2$,
has been shown to yield results in very good agreement with the observations,
and particularly low tensor-to-scalar ratio, while the squared sound speed of
the scalar perturbations is identically equal to 1 and thus the scenario is
free from instabilities. However, this model leads to unitarity
violation which in turn   is  undesirable   if one wished to quantize the
theory.

The consideration of nonminimal derivative couplings of the scalar field to
gravity
is shown to solve the unitarity issue, while still leading to satisfactory
inflationary observables, due to the presence of   a ``gravitational friction''
 that lowers the initial values needed to produce a long slow-roll
period and thus a significant amount of e-folds. Nevertheless, these models
lead to perturbative instabilities and in particular to $c_s^2<0$. Although one
could construct generalized versions of  nonminimal derivative couplings (GNMDC)
that could improve the instability issue, considering for instance a coupling
function of the form
$\frac{\alpha \phi^{\alpha-1}}{M^{\alpha+1}}$, potential problematic behavior
still remains.

In this work we constructed the combined scenario of NMC and of GNMDC, which maintain the advantages of the individual models, but remove their individual disadvantages.
In the combined scenario, a long enough inflationary phase can be
easily achieved, while the initial value of the scalar field  and the scale
of the NMC term   are such that it remains sub-Planckian, a
feature not possible in the single field  NMC scenario.
These attractive features are achieved due to the GNMDC term, that brings about
the  gravitational friction effect  that extends the slow-roll phase,
allowing for lower  initial values of $\phi_*$. Additionally, near the end of
inflation, at the bottom of the potential, when a suitable GNMDC term is chosen, it becomes negligible, while the NMC term  dominates completely. 

To demonstrate this, we have chosen to include two examples, one with a GNMDC of monomial form and one of polynomial form that satisfy the above ansatz (however, other GNMDC forms should also be viable, as long as they become negligible at the end of inflation). In both cases we show that a desirable phenomenology is achieved. At the same time, at the end of inflation
canonical gravity is restored, and the scenario is healed from
$c_s^2$-instabilities due to wild oscillations, and
with no superluminal scalar perturbations that are related to the simple NMDC
case. Finally, another advantage of the present construction is that since
the GNMDC contribution becomes
negligible after inflation ends, the theory can easily pass the recent
LIGO-VIRGO contstraints on the gravitational wave speed
\cite{TheLIGOScientific:2017qsa,Goldstein:2017mmi} (since it is known that
the nonminimal derivative coupling terms are amongst the ones that lead to a
gravitational wave speed different than one).

In summary, the combined scenario leads to inflationary observables in agreement
with observations, it is free from $c_s^2$-instabilities, and it alleviates the
unitarity issue. Hence, it does maintain the advantages of the individual
scenarios without sharing their disadvantages. Thus, inflationary scenarios
with  nonminimal and derivative couplings to gravity combined, may serve as successful
candidates for the description of inflationary dynamics,  and other mechanisms related to inflation, like Primordial Black Hole formation, and deserve further
investigation.

\section*{Acknowledgements}
ENS  would like to acknowledge   the contribution of the
COST Action CA18108 ``Quantum Gravity Phenomenology in the multi-messenger
approach''.

\appendix
\section{Perturbative Analysis of Single Field Inflation}
\label{perturbations}

For the perturbative analysis presented here briefly, we mainly follow \cite{Kobayashi:2011nu,Ema:2015oaa,Tsujikawa:2012mk,Kobayashi:2019}.
The second order action of the curvature perturbation written in the unitary
gauge $\mathcal{R}$, is of the form
\begin{align}\label{pertaction}
	S_{\mathcal{R}}^{(2)}=\int dt d^3x a^3 \left[\mathcal{G}_S
\dot{\mathcal{R}}^2 -\frac{\mathcal{F}_S}{a^2}(\partial \mathcal{R})^2\right]~,
\end{align}
where $a$ is the scale factor, and with the definitions
\begin{align}\label{GsandFsdef}
	\mathcal{G}_S &\equiv
\frac{\Sigma}{\Theta^2}\mathcal{G}_T^2+3\mathcal{G}_T~,   &   \mathcal{F}_S
&\equiv \frac{1}{a}\frac{d}{dt} \left(\frac{a}{\Theta}\mathcal{G}_T^2
\right)-\mathcal{F}_T~.
\end{align}
 $\Sigma, \Theta, \mathcal{G}_T$ and $\mathcal{F}_T$ are functions  that depend
on the Galileon functions and the derivatives of the field (see
\cite{Kobayashi:2019}). Specifically
\begin{align}\label{Gtdefinition}
	\mathcal{G}_T\equiv 2 \left[ G_4-2X G_{4X}-X \left(H\df
G_{5X}-G_{5\phi}\right)\right]~,
\end{align}
\begin{align}\label{Ftdefinition}
\!\!\!\!\!\!\!\!\!\!\!\!\!\!\!\!\!\!\!\!\!\!\!\!\!\!\!\!\!\!\!\!\!\!\!\!\!\!\!\!
\!\!\!\!\!\!\!	\mathcal{F}_T\equiv 2 \left[ G_4-X
\left(\ddot{\phi}G_{5X}+G_{5\phi}\right)\right]~,
\end{align}
\begin{align}\label{Thetadefinition}
	\Theta &\equiv -\df X G_{3X}+2H\left(G_4-4 X G_{4X}-4X^2 G_{4XX}\right)\nonumber \\
	&+\df
\left( G_{4\phi}+2X G_{4\phi X}\right)-H^2\df \left(5XG_{5X}+2X^2
G_{5XX}\right)\notag \\
	&+2HX \left(3 G_{5\phi}+2XG_{5\phi X}\right)~,
\end{align}
\begin{align}\label{Sigmadefinition}
	\Sigma \equiv& X \left(G_{2X}+2X G_{2XX}\right)+6 H\df X \left( 2 G_{3X}+X
G_{3XX}\right)\nonumber \\
&
\!
-2X \left( G_{3\phi}+X G_{3\phi X}\right)-6H^2 G_4 \notag \\
	&\!+6\left[ H^2 \left( 7X G_{4X}+16 X^2 G_{4XX}+4X^3
G_{4XXX}\right)\right.\nonumber \\
	&\!\left.-H\df
\left(G_{4\phi}+5X G_{4\phi X}+2 X^2 G_{4\phi XX}\right)\right] \notag \\
	&\!+H^3 \df X\left(30G_{5X}+26 X G_{5XX}+4 X^2 G_{5XXX} \right)\nonumber \\
	&\!-6 H^2 X \left(
6 G_{5\phi}+9 X G_{5\phi X}+2X^2 G_{5\phi XX}\right)~.
\end{align}
Then, with specified Galileon functions, one can calculate the squared sound
speed of the scalar and tensorial perturbations via the formulas
\begin{align} \label{sspddef}
	c_s^2 &\equiv \frac{\mathcal{F}_S}{\mathcal{G}_S}~,   &   c_T^2 &\equiv
\frac{\mathcal{F}_T}{\mathcal{G}_T}~.
\end{align}
These quantities have to be positive in order to avoid gradient instabilities,
i.e. exponential growth of perturbation modes. On the same footing, to ensure
that the kinetic terms are positive and no ghost instabilities appear, the
constraints
\begin{align}\label{consghost}
	\mathcal{G}_S>0~,~~~~   &   \mathcal{G}_T>0~,
\end{align}
must   hold, too.
For instance, in the  NMC scenario, using equations
\eqref{Fried00NMC},\eqref{Fried11NMC} and \eqref{KGinNMC},
it is straightforward to prove that $c_s^2=1$, hence no corresponding
instabilities occur.

Furthermore, one can show that in order to calculate the Power
Spectrum of the scalar and tensorial perturbations, one simply needs to
calculate \cite{Mukhanov:1992}
\begin{align}\label{PsandPtdef}
	\mathcal{P}_\mathcal{R} =
\frac{\mathcal{G}_S^{1/2}}{2\mathcal{F}_S^{3/2}}\frac{H^2}{4\pi^2}~, \  &
\mathcal{P}_\mathcal{T} =
\frac{8\mathcal{G}_T^{1/2}}{\mathcal{F}_T^{3/2}}\frac{H^2}{4\pi^2}~,
\end{align}
which are evaluated at the horizon crossing. Their quotient is the
tensor-to-scalar ratio $r$, namely
\begin{align} \label{rdef}
	r \equiv \frac{\mathcal{P}_\mathcal{T}}{\mathcal{P}_\mathcal{R}}~.
\end{align}
Furthermore,  we   introduce the scalar spectral index, $n_s$, expressing
the change of
the logarithm of the scalar power spectrum per logarithmic interval $k$, via
the
relation
\begin{align}
	1-n_s \equiv-\frac{d \ln\mathcal{P}_{\mathcal{R}}}{d \ln k}\Big|_{k=aH}~,
	\label{nsdefgene}
\end{align}
and likewise, the tensorial spectral index
\begin{align}
	n_t \equiv-\frac{d \ln\mathcal{P}_{\mathcal{T}}}{d \ln k}\Big|_{k=aH}~.
\end{align}
The tensor-to-scalar ratio and the tensor tilt are related via what is called
the consistency condition and in standard  single-field inflation it takes the
form $r \approx -8 n_t$, nevertheless in modified  scenarios its form can be
non-standard.
In summary, the observables of an inflationary model are $r, n_s$ and
$\mathcal{P}_\mathcal{R}$.

\end{document}